\documentclass[english]{article}
\usepackage[T1]{fontenc}
\usepackage[latin9]{inputenc}
\usepackage{float}
\usepackage{mathrsfs}
\usepackage{amstext}
\usepackage{graphicx}
\usepackage{setspace}
\makeatletter
\usepackage[numbers, sort&compress]{natbib}

\makeatother

\usepackage{babel}
\begin{document}

\title{Quantum Entanglement Establishment between two Strangers}

\author{Tzonelih Hwang\thanks{Corresponding author\protect \\
hwangtl@ismail.csie.ncku.edu.tw\protect \\
Department of Computer Science and Information Engineering, National
Cheng Kung University, No. 1, University Rd., Tainan City, 701, Taiwan,
R.O.C.}, Tzu-Han Lin, and Shih-Hung Kao}
\maketitle
\begin{abstract}
This paper presents the first quantum entanglement establishment scheme
for strangers who neither pre-share any secret nor have any authenticated
classical channel between them. The proposed protocol requires only
the help of two almost dishonest third parties (TPs) to achieve the
goal. The security analyses indicate that the proposed protocol is
secure against not only an external eavesdropper's attack, but also
the TP's attack.


\end{abstract}

\section{Introduction}

Quantum entanglement, one of the most attractive physical phenomena,
has been widely researched in recent years. Quantum entanglement provides
a \textquotedbl{}spooky relation at a distance,\textquotedbl{} which
allows two or more participants who share entangled quantum states
to have correlated information. Based on the concept of quantum entanglement,
various quantum cryptographic protocols are possible. For example,
quantum key distribution allows two remote participants to share a
secure key \cite{E91}; quantum teleportation \textquotedbl{}sends\textquotedbl{}
quanta to a remote location without any physical photon transmission
\cite{Telep}; quantum dense-coding communication allows one to transmit
two-bit information via a one-bit quantum transmission; and quantum
blind computation \cite{QBlindComp} allows a user to perform quantum
computations with the help of a quantum server, without revealing
the intended computations. In addition, quantum secret sharing \cite{QSS},
quantum state sharing \cite{QStateS}, quantum remote state preparation
\cite{QRSP}, quantum signature \cite{QSignature}, quantum private
comparison \cite{QPC-1,QPC-2}, etc., are all possible because of
shared quantum entanglement states. Research has shown that if shared
quantum entanglements are in incorrect states or are interrupted by
malicious users during the entanglement establishment process, then
incorrect results may occur, and the protocol is considered to be
insecure \cite{E91,TP4-attack,WrongEntangle-1,WrongEntangle-2,CE-1,CE-2,CE-3,CE-4}.
Accordingly, assurance of security and correctness during the establishment
of entanglement becomes an imperative issue in quantum cryptography.

The problem with the establishment of entanglement has been most often
treated in two ways. The first is to simply assume that the entanglement
is pre-shared by the participants \cite{Telep}. The second--which
is also our focus here--describes the entanglement establishment procedure
in detail \cite{E91}. For this approach, one often assumes the existence
of an \textbf{authentication classical channel} between two users,
which can be used to discuss the correctness of the shared entangled
states. For example, if Alice wants to share an entanglement with
Bob, Alice will generate a series of entangled quantum states, which
include multiple quantum particles, and transmit the entangled particles
to Bob. Then Alice and Bob choose some entangled states for public
discussion: they respectively measure the selected entangled states
and compare the measurement result via the authenticated classical
channel. If the comparison result is accepted, then Alice and Bob
believe that the entanglement is well established. 

However, to share an authenticated classical channel, the implication
is that Alice and Bob should know each other beforehand. What if Alice
and Bob are strangers--i.e., they did not meet each other beforehand?
In that case, Alice and Bob might have to look for another client--say,
Charlie--as a third party (TP) \cite{TP1,TP2,TP3}, who respectively
shares a quantum channel and an authenticated classical channel with
them and eventually can help them share an entanglement. The trustworthiness
of this TP has been an interesting topic in quantum cryptography.
The issue surrounds the usefulness of the constructed protocol in
practice. The ideal case assumes the existence of a completely trusted
TP who always executes the protocol loyally and never reveals the
important information of the users. This case is conceptually the
same as assuming the existence of an authenticated classical channel
between two involved users. In these cases, the TP is assumed to be
a \textbf{semi-honest} agent who will loyally execute the protocol,
but may try to steal Alice and Bob's secret using passive attacks.
The semi-honest TP will passively collect the classical information
exchanged between Alice and Bob and try to reveal their secrets from
this information \cite{QPC-1,QPC-2}.

In the other more practical cases, the TP is assumed to be \textbf{almost
dishonest} and may deviate from the normal procedure of the protocol
to reveal the participants' secret information except acting in collusion
with the clients. That is, the TP not only can passively collect useful
information but also can actively perform any attack on the protocol
except conspiring with the participant. In this case, however, Alice
and Bob, who are strangers and thus do not directly share an authenticated
channel, may not be able to detect and avoid the TP's attack, and
their secret information might thus be leaked to the TP \cite{TP4-attack}.
In this regard, can we also develop a secure protocol for a pair of
strangers to share entanglement under the help of almost dishonest
TPs? This is the question addressed by this work. 

In order to do that, we assume the existence of two non-communicating
TPs. That is, Alice and Bob attempt to find two TPs from the group
of clients who simultaneously share both the quantum channels and
authenticated classical channels with them. Because the chosen TPs
are non-communicating, they do not know who will be the counter party--e.g.,
the other TP--of the current scenario. This also increases the difficulty
of collusion between both TPs. That is, the TPs are non-communicating
and do not know each other, and hence it is more difficult for them
to act in collusion.

If we let the first almost dishonest TP generate entangled quantum
states for Alice and Bob, and the second almost dishonest TP helps
Alice and Bob check the correctness of the shared entanglement. Moreover,
let each TP watch the other TP's malicious behavior; then the entanglement
could thus be established between Alice and Bob securely and correctly.

It should be noted here that the power of the conventional TP is now
divided into two parts, which is similar to the idea of secret sharing,
in which a top secret is divided into several shadows and only when
a sufficient amount $t$ of shadows is collected can the top secret
be derived. Fewer than $t$ shadows will never reveal the top secret.
In our case, owing to this power separation of TPs and due to the
assumption of non-colluding TPs, the trustworthiness of TPs in our
protocol can be demoted from semi-honest to almost dishonest.

Now, because the TPs are non-communicating and almost dishonest, other
clients in the protocol can serve well as TPs as long as they simultaneously
share both quantum channels and authenticated classical channels with
both clients who want to establish entanglement in the protocol. In
other words, within our protocol, the only involved roles are the
clients themselves. Among them, some connect with quantum or authenticated
classical channels whereas others are strangers who do not have any
direct connection. Whoever wants to establish entanglement with the
other, has to identify two other clients serving as their TPs, who
can directly communicate with both users in both the quantum channel
and the authenticated classical channel. (see also Fig. 1)

The rest of this paper is organized as follows: Section 2 describes
the proposed environment and the quantum entanglement establishment
protocol between strangers; a quantum communication protocol is also
presented as an example. Section 3 provides the security analyses
of the proposed protocol. Finally, the conclusions are presented in
Section 4.

\section{Proposed Protocol}

This section introduces the proposed protocol. First, the environment
with two TPs is introduced in Section 2.1; then, the proposed quantum
entanglement establishment protocol is presented in Section 2.2. Finally,
as an example, a quantum secure direct communication protocol based
on the proposed entanglement establishment is described in Section
2.3.

\subsection{Environment}

This section describes the proposed environment and its security requirement.
The environment, including two participants, Alice and Bob, and two
TPs, TP1 and TP2, is described as follows. (see also Fig. 1, in which
the dotted lines denote the quantum channels and the solid lines denote
the authenticated classical channels.)

\begin{figure}
\begin{centering}
\includegraphics[bb=74bp 200bp 950bp 568bp,clip,scale=0.4]{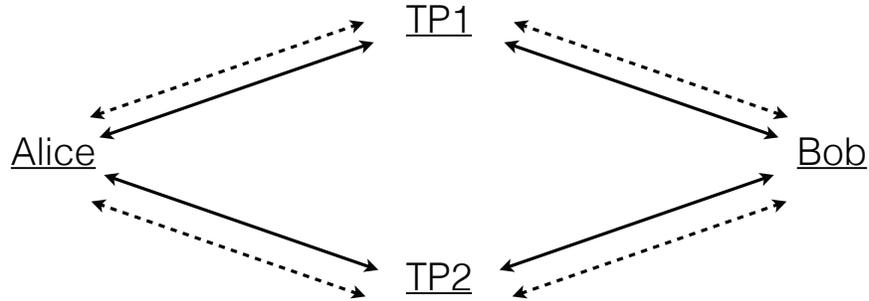}
\par\end{centering}

\caption{The Proposed Environment}
\end{figure}

\begin{enumerate}
\item Alice (Bob) shares authenticated classical channels and quantum channels
with two TPs, respectively. Note that Alice and Bob do not have any
authenticated channel directly connected to each other.
\item The transmitted information on the authenticated classical channel
is public, but the receiver can verify its integrity and originality.
\item TPs are almost dishonest in the sense that they can perform any possible
attacks except conspiring with Alice, Bob, or the other TP.
\item In our proposed environment, if a TP cannot successfully attack the
shared entanglement, then the TP will not announce the fake results
during the public discussion.
\item Each TP is designed to prevent the other TP from attacks; hence, both
TPs can be almost dishonest.
\item An external attacker, Eve, may try to perform any attack to disturb,
forge, or eavesdrop on the state of Alice and Bob's shared entanglement.
\end{enumerate}

\subsection{Proposed Quantum Entanglement Establishment}

This section comprises a quantum entanglement establishment protocol
for the proposed environment. The proposed protocol allows the sender,
Alice to share an entangled state with the stranger Bob. In the proposed
protocol, though the quantum signals are generated by TP1, Alice and
Bob can determine whether TP1 performs any attack on the quantum signals
with the help of TP2. The proposed protocol proceeds as follows (see
Fig. 2):

\begin{figure}[H]
\begin{centering}
\includegraphics[bb=0bp 150bp 1024bp 710bp,clip,scale=0.35]{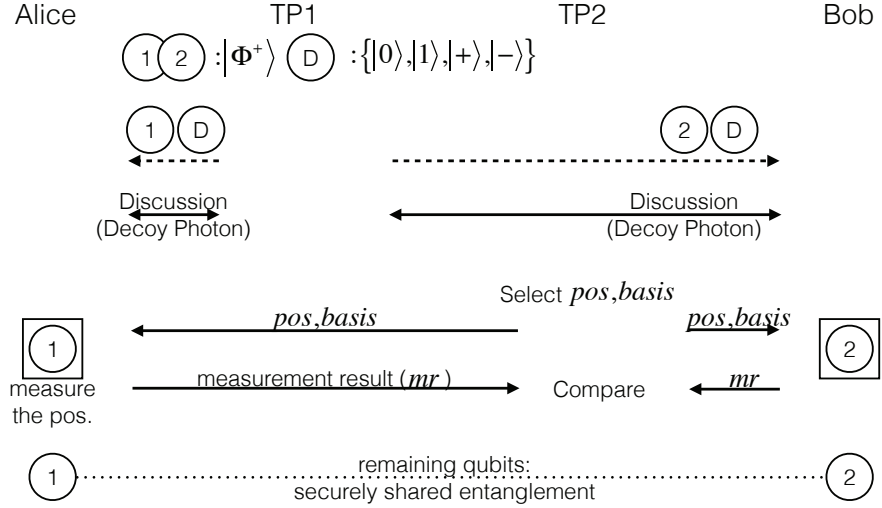}
\par\end{centering}

\caption{The Proposed Entanglement Establishment.}
\end{figure}

\begin{description}
\item [{(Step1)}] TP1 generates a sequence of EPR entangled states, $\left|\Phi^{+}\right\rangle _{12}=\frac{1}{\sqrt{2}}\left(\left|00\right\rangle \right.$
$\left.+\left|11\right\rangle \right)_{12}$, where the subscripts
1 and 2 denote respectively the first and the second qubits. Let $Q_{1}$
($Q_{2}$) denotes the particle sequence includes all the first (second)
qubit of each EPR state in order. TP1 then inserts enough amount of
decoy photons \cite{DECOY-3,DECOY-1,DECOY-2} randomly chosen from
the four states: $\left\{ \left|0\right\rangle ,\left|1\right\rangle ,\right.$
$\left.\left|+\right\rangle =\frac{1}{\sqrt{2}}\left(\left|0\right\rangle +\left|1\right\rangle \right),\left|-\right\rangle =\frac{1}{\sqrt{2}}\left(\left|0\right\rangle -\left|1\right\rangle \right)\right\} $
into $Q_{1}$ ($Q_{2}$) to form a new sequence $S_{1}$ ($S_{2}$.)
TP1 sends the sequence $S_{1}$ to Alice, and $S_{2}$ to Bob, respectively.
\item [{(Step2)}] Once Alice receives the quantum sequence $S_{1}$ from
TP1, she sends an acknowledgement to TP1 via the authenticated classical
channel. TP1 and Alice then will publicly discuss the decoy photons
for the eavesdropping detection. TP1 informs the position and the
basis of each decoy photon to Alice. Alice measures these decoy photons,
and then sends the measurement results to TP1. By comparing the initial
states and the measurement results, TP1 can detect the existence of
eavesdroppers. Similarly, Bob will also publicly discuss the decoy
photons in $S_{2}$ with TP1. According to the quantum cryptographic
protocol which will be executed after entanglement establishment,
if Alice will send the received particles out later, Alice has to
use the photon number splitter (PNS) and the wavelength filter to
check if Trojan Horse attacks exist in the protocol \cite{THA-2,THA-1,THA-3,THA-4}.
(The detailed analyses of the Trojan Horse attacks will be given in
Section 3.2)
\item [{(Step3)}] If the quantum transmissions are free from the eavesdroppers
and the Trojan Horse attacks, Alice and Bob can remove the decoy photons
and recover the sequences $Q_{1}$ and $Q_{2}$. They then will discuss
the entanglement of the shared EPR states via the help of TP2. TP2
randomly selects the position and basis ($X$ basis or $Z$ basis)
for each photon to be checked and announces the positions and bases
to Alice and Bob. Alice and Bob then measure the selected particles
in $Q_{1}$ ($Q_{2}$) with the bases chosen by TP2, and sends the
measurement results to TP2. Because

\begin{equation}
\begin{array}{ccc}
\left|\Phi^{+}\right\rangle  & = & \frac{1}{\sqrt{2}}\left(\left|00\right\rangle +\left|11\right\rangle \right)\\
 & = & \frac{1}{\sqrt{2}}\left(\left|++\right\rangle +\left|--\right\rangle \right),
\end{array}
\end{equation}
\\
TP2 can compare the measurement results from Alice to Bob to determine
if Alice and Bob's qubits are in $\left|\Phi^{+}\right\rangle $.

\item [{(Step4)}] If the entanglement correlations between Alice's and
Bob's qubits are correct, Alice (Bob) will remove the measured qubits
selected by TP2 from $Q_{1}$ ($Q_{2}$,) and have a new sequence,
$Q_{1}^{\prime}$ ($Q_{2}^{\prime}$.) The entanglement establishment
between Alice and Bob is completed.
\end{description}
It should be noted that the proposed entanglement establishment scheme
can also be extended to a multi-participant scenario. Suppose that
the participants--Alice, Bob, Charlie, ..., and Zack, who are strangers
to one another--want to share an entanglement. They can look for two
almost dishonest TPs who share quantum channels and authenticated
classical channels with all participants (see Fig. 3). TP1 can generate
the entanglement and securely distribute the particles to them, which
are the same as Step 1 and Step 2; then, similar to Step 3, TP2 selects
the measurement bases and positions for each participant and compares
the measurement results returned from every participant. Hence, TP2
can help the participants check the correctness of the entanglement
they shared. Finally, the participants can share an entanglement.

\begin{figure}
\begin{centering}
\includegraphics[bb=74bp 200bp 950bp 568bp,clip,scale=0.4]{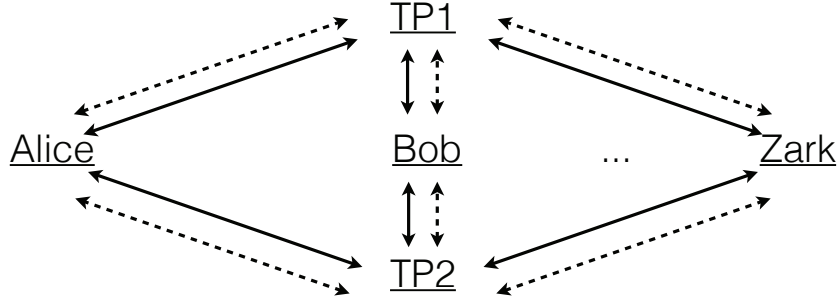}
\par\end{centering}

\caption{The Multiparty Environment}
\end{figure}

\subsection{Quantum Secure Direct Communication}

Section 2.2 describes the processes of the proposed quantum entanglement
establishment protocol. In this section, we show how a quantum secure
direct communication (QSDC) protocol can be constructed based on the
entanglement establishment protocol given in Section 2.2. In a QSDC
protocol, a sender can send secret messages to a receiver without
any pre-shared key between them, and they do not require any transmission
of classical information except for the eavesdropping detection. Here
we assume that Alice wants to send a two-bit message to Bob (see Fig.
4).

\begin{figure}
\begin{centering}
\includegraphics[bb=0bp 150bp 1024bp 710bp,clip,scale=0.35]{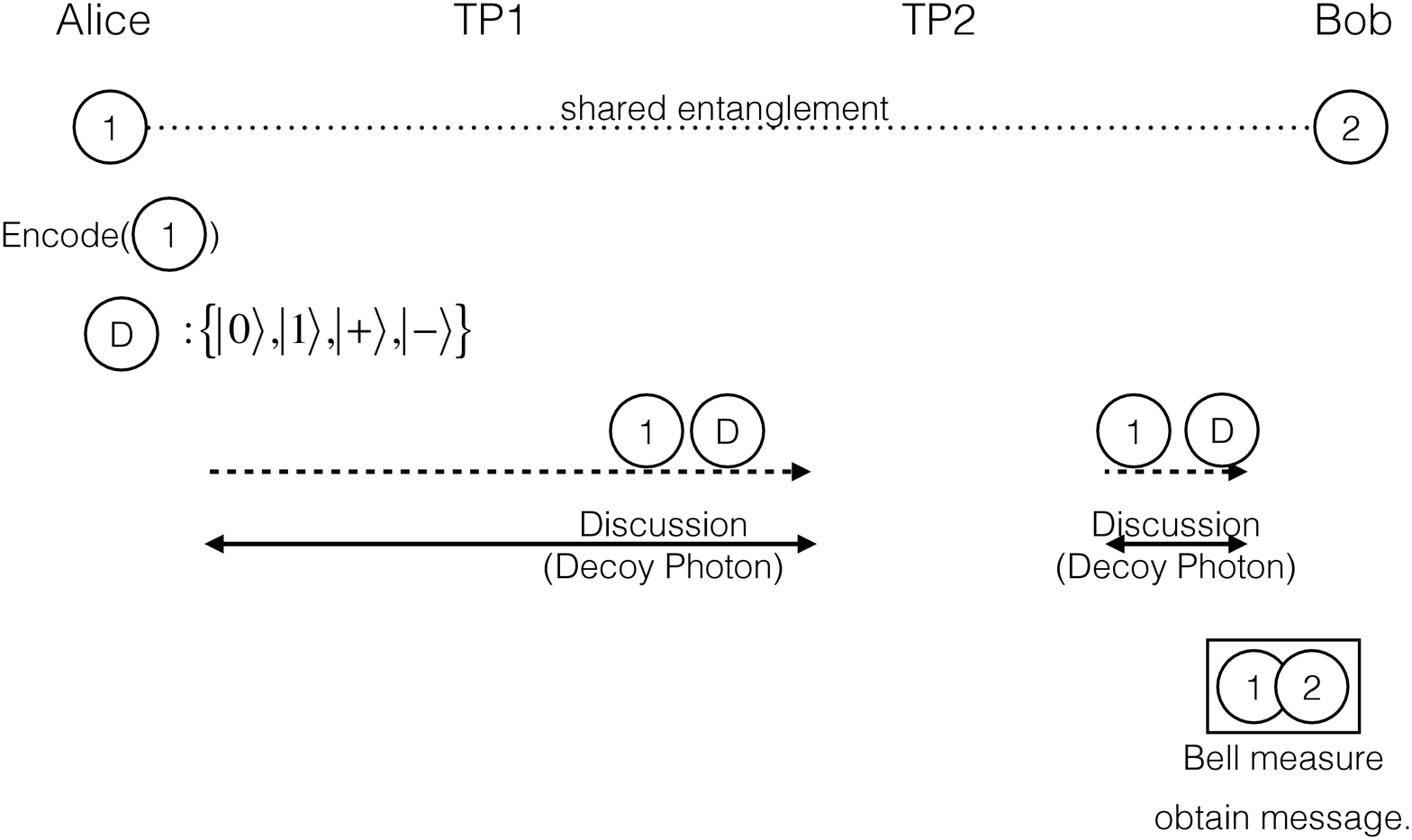}
\par\end{centering}

\caption{The Quantum Direct Secure Communication based on the Proposed Entanglement
Establishment.}
\end{figure}

\begin{description}
\item [{(Step1\textasciitilde{}4)}] These steps are the same as those mentioned
in Section 2.2. 
\item [{(Step5)}] To transmit her secret message, Alice applies dense coding
on her photons by performing the unitary operation on each qubit of
$Q_{1}^{\prime}$ obtained in Section 2.2 according to her two-bit
messages. If the two-bit message is 00, she will perform $I=\left|0\right\rangle \left\langle 0\right|+\left|1\right\rangle \left\langle 1\right|$;
if the two-bit message is 01, she will perform $\sigma_{z}=\left|0\right\rangle \left\langle 0\right|-\left|1\right\rangle \left\langle 1\right|$;
if the two-bit message is 10, she will perform $\sigma_{x}=\left|0\right\rangle \left\langle 1\right|+\left|1\right\rangle \left\langle 0\right|$;
otherwise, she will perform $i\sigma_{y}=\left|0\right\rangle \left\langle 1\right|-\left|1\right\rangle \left\langle 0\right|$.
After the encoding, Alice generates decoy photons as in Step 1 of
Section 2.2 by TP1, and inserts them to $Q_{1}^{\prime}$ to form
a new sequence $S_{1}^{\prime}$, which is then transmitted to TP2.
Upon receiving $S_{1}^{\prime}$, TP2 publicly discusses the decoy
photons with Alice as in Step 2. If there are eavesdroppers detected,
they will abort the protocol and return to Step 1.
\item [{(Step6)}] TP2 removes the decoy photons from $S_{1}^{\prime}$,
and inserts new decoy photons into the particle sequence to form $S_{1}^{\prime\prime}$,
which is then transmitted to Bob.
\item [{(Step7)}] Bob and TP2 again discuss the decoy photons for detecting
the eavesdroppers. If the quantum transmission between TP2 and Bob
is secure, Bob can remove the decoy photons and obtain the particle
sequence $Q_{1}^{\prime}$. Bob then performs Bell measurement (EPR
measurement) on every pair of qubits respectively from $Q_{1}^{\prime}$
and $Q_{2}^{\prime}$. According to the measurement results, Bob can
obtain Alice's secret message. (See Eq. (2))

\begin{equation}
\begin{array}{ccccc}
I\left|\Phi^{+}\right\rangle  & = & \frac{1}{\sqrt{2}}\left(\left|00\right\rangle +\left|11\right\rangle \right) & = & \left|\Phi^{+}\right\rangle \\
\sigma_{z}\left|\Phi^{+}\right\rangle  & = & \frac{1}{\sqrt{2}}\left(\left|00\right\rangle -\left|11\right\rangle \right) & = & \left|\Phi^{-}\right\rangle \\
\sigma_{x}\left|\Phi^{+}\right\rangle  & = & \frac{1}{\sqrt{2}}\left(\left|01\right\rangle +\left|10\right\rangle \right) & = & \left|\Psi^{+}\right\rangle \\
i\sigma_{y}\left|\Phi^{+}\right\rangle  & = & \frac{1}{\sqrt{2}}\left(\left|01\right\rangle -\left|10\right\rangle \right) & = & \left|\Psi^{-}\right\rangle 
\end{array}
\end{equation}

\end{description}

\section{Security Analyses}

This session analyzes the security of the proposed protocols. First,
in Section 3.1, the security of the entanglement establishment protocol
is analyzed. The security of the QSDC protocol is analyzed in Section
3.2. The formal security proof using the random oracle model is described
in the appendix.

\subsection{Security of the Entanglement Establishment Protocol}

For an entanglement establishment protocol, the attacker may try to
obtain Alice's entanglement qubits. Hence, the attacker can share
a quantum entanglement with Bob. Here, three possible attack strategies
will be discussed: the entangle-and-measure attack, the intercept-and-resend
attack, and the entanglement swapping attack. The analyses indicate
that the external attacker and the TPs in the protocol cannot successfully
obtain the entanglement shared between Alice and Bob without being
detected.

\subsubsection*{The Entangle-and-measure Attack}

When TP1 sends $S_{1}$ to Alice in Step 1, the external eavesdropper,
Eve, may perform the entangle-and-measure attack \cite{EMA-1,EMA-3,EMA-2}
to steal the transmitted qubits in $S_{1}$. Because $S_{1}$ contains
TP1's decoy photons, to avoid being detected, Eve will try to obtain
the states of the decoy photons. For each qubit, $q_{1}$, in $S_{1}$,
Eve prepares an ancillary qubit in an arbitrary known state $q_{e}=\left|E\right\rangle $,
and performs her attack operation $U$ on $q_{1}$ and $q_{e}$. The
result of Eve's operation is as follows:

\begin{equation}
\begin{array}{ccc}
U\left|0\right\rangle _{1}\left|E\right\rangle _{e} & = & a\left|0\right\rangle _{\text{1}}\left|e_{00}\right\rangle _{e}+b\left|1\right\rangle _{1}\left|e_{01}\right\rangle _{e}\\
U\left|1\right\rangle _{1}\left|E\right\rangle _{e} & = & c\left|0\right\rangle _{1}\left|e_{10}\right\rangle _{e}+d\left|1\right\rangle _{1}\left|e_{11}\right\rangle _{e}\\
U\left|+\right\rangle _{1}\left|E\right\rangle _{e} & = & \frac{1}{2}\left[\begin{array}{l}
\left|+\right\rangle _{1}\left(a\left|e_{00}\right\rangle _{e}+b\left|e_{01}\right\rangle _{e}+c\left|e_{10}\right\rangle _{e}+d\left|e_{11}\right\rangle _{e}\right)+\\
\left|-\right\rangle _{1}\left(a\left|e_{00}\right\rangle _{e}-b\left|e_{01}\right\rangle _{e}+c\left|e_{10}\right\rangle _{e}-d\left|e_{11}\right\rangle _{e}\right)
\end{array}\right]\\
U\left|-\right\rangle _{1}\left|E\right\rangle _{e} & = & \frac{1}{2}\left[\begin{array}{l}
\left|+\right\rangle _{1}\left(a\left|e_{00}\right\rangle _{e}+b\left|e_{01}\right\rangle _{e}-c\left|e_{10}\right\rangle _{e}-d\left|e_{11}\right\rangle _{e}\right)+\\
\left|-\right\rangle _{1}\left(a\left|e_{00}\right\rangle _{e}-b\left|e_{01}\right\rangle _{e}-c\left|e_{10}\right\rangle _{e}+d\left|e_{11}\right\rangle _{e}\right)
\end{array}\right],
\end{array}
\end{equation}
\\
where $\left|e_{00}\right\rangle $, $\left|e_{01}\right\rangle $,
$\left|e_{10}\right\rangle $, and $\left|e_{11}\right\rangle $ are
four states which Eve can distinguish, and $\left|a^{2}\right|+\left|b^{2}\right|=\left|c^{2}\right|+\left|d^{2}\right|=1$.

To pass the eavesdropping detection, Eve sets $b=c=0$ and $\left(a\left|e_{00}\right\rangle _{e}+b\left|e_{01}\right\rangle _{e}\right.$
$\left.+c\left|e_{10}\right\rangle _{e}+d\left|e_{11}\right\rangle _{e}\right)$
$=\left(a\left|e_{00}\right\rangle _{e}+b\left|e_{01}\right\rangle _{e}-c\left|e_{10}\right\rangle _{e}-d\left|e_{11}\right\rangle _{e}\right)=\overrightarrow{0}$.
Eve's operation thus will not change the state of $q_{1}$, and Eve
can successfully pass the eavesdropping detection. However, $b=c=0$
implies $\left(a\left|e_{00}\right\rangle _{e}-d\left|e_{11}\right\rangle _{e}\right)=\overrightarrow{0}$,
that is, $a\left|e_{00}\right\rangle _{e}=d\left|e_{11}\right\rangle _{e}$.
In this case, Eve cannot distinguish $\left|e_{00}\right\rangle $
and $\left|e_{11}\right\rangle $, and she cannot obtain the information
in $q_{1}$. If Eve wants to distinguish $a\left|e_{00}\right\rangle _{e}$
from $d\left|e_{11}\right\rangle _{e}$, her operation, $U$, will
change the state of $q_{1}$, which will cause her attack to be detected
by TP1 and Alice.

Generally, if Eve want to pass the eavesdropping check, she cannot
get any information. If Eve tries to reveal the whole information
from a qubit, she will change the state of the qubit, and eventually
be detected in the public discussion.

\subsubsection*{The Intercept-and-resend Attack}

TP2 may perform the intercept-and-resend attack when TP1 sends $S_{1}$
to Alice. TP2 intercepts all qubits in $S_{1}$, and generates a sequence
of fake photons, which are sent to Alice. If TP2 can pass the eavesdropping
detection process, TP2 then could successfully share an entanglement
with Bob. However, the decoy photons inserted in $S_{1}$ are generated
by TP1. Because the positions and the bases of the decoy photons are
unknown to TP2, TP2 is unable to exactly generate the same decoy photons
as TP1 did. Hence, TP2's fake photons will cause errors in the public
discussion between Alice and TP1 with the probability $1-\left(75\%\right)^{n}$
\cite{DECOY-1,DECOY-2,DECOY-3}, where $n$ is the number of decoy
photons. If $n$ is large enough, the probability will be close to
1.

\subsubsection*{The Entanglement Swapping Attack}

TP1, who generates the EPR states for Alice and Bob, may also try
to perform the entanglement swapping attack \cite{TP4-attack} to
obtain Alice's secret message. In Step 1, instead of generating one
EPR pair and distributing these two particles to Alice and Bob, respectively,
TP1 generates two EPR pairs, namely $\left|\Phi^{+}\right\rangle _{T1,T2}$
and $\left|\Phi^{+}\right\rangle _{T3,T4}$. TP then distributes $q_{T1}$,
the first particle of the first EPR state, to Alice, and $q_{T3}$,
the first particle of the second EPR state, to Bob. Because all the
decoy photons are generated by TP1, TP1 can successfully pass the
eavesdropping check of decoy photons in Step 2. If the entanglement
correlation check in Step 3 can be passed, in Step 5, Alice will send
the encoded particle, $q_{T1}$ to TP2. TP1 can intercept them, remove
the decoy photons according to Alice and TP2's public communication,
and perform an EPR measurement on the qubit pair $q_{T1}$ and $q_{T2}$.
According to the measurement result, TP1 can obtain Alice's secret
message.

But the fact is , when Alice, TP2, and Bob discuss the entanglement
of the shared EPR states in Step 3, for each discussed position, TP1
can measure the qubit pair $q_{T2}$ and $q_{T4}$. The qubits $q_{T1}$
and $q_{T3}$ then will be in one of four EPR states, $\left\{ \left|\Phi^{+}\right\rangle ,\left|\Phi^{-}\right\rangle ,\left|\Psi^{+}\right\rangle ,\left|\Psi^{-}\right\rangle \right\} $,
which is also known by TP1. Because these two particles held respectively
by Alice and Bob are still in EPR state, they cannot detect that TP1
generated two EPRs rather than one. However, the above situation happens
only when TP1 is allowed to generate variable EPR states. In the proposed
protocol, however, TP1 is only allowed to generate $\left|\Phi^{+}\right\rangle $,
if he/she performs the above attack, the EPR state shared by Alice
and Bob in public discussion will be in one of $\left\{ \left|\Phi^{+}\right\rangle ,\left|\Phi^{-}\right\rangle ,\left|\Psi^{+}\right\rangle ,\left|\Psi^{-}\right\rangle \right\} $,
rather than in $\left|\Phi^{+}\right\rangle $ as in normal situation.
For example, if the shared state is $\left|\Psi^{-}\right\rangle =\frac{1}{\sqrt{2}}\left(\left|01\right\rangle -\left|10\right\rangle \right)$,
Alice's and Bob's Z-basis measurement will be $\left|0\right\rangle $
and $\left|1\right\rangle $ ($\left|1\right\rangle $ and $\left|0\right\rangle $,)
whereas the legal measurement results are $\left|0\right\rangle $
and $\left|0\right\rangle $ ($\left|1\right\rangle $ and $\left|1\right\rangle $).
TP1's attack thus will be detected by TP2. The proposed protocol is
thus secure against TP1's entanglement swapping attack.

\subsection{Security of QSDC Protocol}

The above analyses denote the security of the proposed entanglement
establishment scheme. The following analyses focus on the security
of the QSDC protocol. Four special attacks--the Trojan Horse attacks,
the correlation-elicitation (CE) attack, the dense coding attack,
and the modification attack--will be respectively analyzed. We also
indicate that the QSDC protocol satisfies the Deng-Long criteria,
a security requirement for quantum communication protocols.

\subsubsection*{The Trojan Horse Attacks}

Eve (TP1, TP2) may perform the Trojan Horse attacks to reveal Alice's
message. When TP1 sends $S_{1}$ to Alice, she can insert her own
particle sequence into $S_{1}$ by adopting the invisible-photon attack
\cite{THA-1} or the delay-photon attack \cite{THA-2} strategy. When
Alice sends out $S_{1}^{\prime}$ to TP2 after her encoding, Eve can
retrieve her particles, and then obtains Alice\textquoteright s secret.
However, because Alice has set a wavelength filter and PNS, Eve\textquoteright s
particles can be detected by these devices. If illegal particles are
detected, Alice and TP1 will drop these transmitted particles and
restart the protocol. Hence, Eve cannot obtain any information about
Alice\textquoteright s secret.

\subsubsection*{The Correlation-elicitation Attack}

The almost dishonest TP2 may try to steal Alice's secret by performing
the correlation-elicitation (CE) attack \cite{CE-2,CE-1,CE-3,CE-4}.
When the second qubit of the EPR state generated by TP1 (denoted as
$q_{2}$) is transmitted to Bob in Step 1, TP2 intercepts it, and
generates an ancillary photon $q_{e}=\left|0\right\rangle $. TP2
then performs the first controlled-NOT (CNOT) operation on $q_{2}$
and $q_{e}$, where $q_{2}$ is the control bit, and $q_{e}$ is the
target bit. As a result, the two-particle EPR state and the ancillary
photon can be described as follows:

\begin{equation}
CNOT_{2e}\left|\Phi^{+}\right\rangle _{12}\otimes\left|0\right\rangle _{e}=\frac{1}{\sqrt{2}}\left(\left|000\right\rangle +\left|111\right\rangle \right),
\end{equation}
\\
where $\otimes$ denotes the tensor product operation. TP2 then resends
$q_{2}$ to Bob. When Alice sends the encoded first qubit, $q_{1}$,
of the EPR state to TP2 in Step 5, TP2 performs the second CNOT operation
on $q_{1}$ and $q_{e}$, where $q_{1}$ is the control bit, and $q_{e}$
is the target bit. Due to Alice's encoding operation, the state of
$q_{1}$ and $q_{2}$ becomes one of $\left|\Phi^{+}\right\rangle $,
$\left|\Phi^{-}\right\rangle $, $\left|\Psi^{+}\right\rangle $,
and $\left|\Psi^{-}\right\rangle $ (see Eq. (2).) The four possible
states after the second CNOT operation are as follows:

\begin{equation}
\begin{array}{lll}
CNOT_{1e}CNOT_{2e}\left|\Phi^{+}\right\rangle _{12}\otimes\left|0\right\rangle _{e} & = & \frac{1}{\sqrt{2}}\left(\left|00\right\rangle +\left|11\right\rangle \right)_{12}\otimes\left|0\right\rangle _{e}\\
CNOT_{1e}CNOT_{2e}\left|\Phi^{-}\right\rangle _{12}\otimes\left|0\right\rangle _{e} & = & \frac{1}{\sqrt{2}}\left(\left|00\right\rangle -\left|11\right\rangle \right)_{12}\otimes\left|0\right\rangle _{e}\\
CNOT_{1e}CNOT_{2e}\left|\Psi^{+}\right\rangle _{12}\otimes\left|0\right\rangle _{e} & = & \frac{1}{\sqrt{2}}\left(\left|01\right\rangle +\left|10\right\rangle \right)_{12}\otimes\left|1\right\rangle _{e}\\
CNOT_{1e}CNOT_{2e}\left|\Psi^{-}\right\rangle _{12}\otimes\left|0\right\rangle _{e} & = & \frac{1}{\sqrt{2}}\left(\left|01\right\rangle -\left|10\right\rangle \right)_{12}\otimes\left|1\right\rangle _{e}
\end{array}
\end{equation}

TP2 is now able to obtain Alice's partial secret according to the
Z-basis measurement result of $q_{e}$. According to Eq. (5,) if the
measurement result of $q_{e}$ is $\left|0\right\rangle $, TP2 knows
that the state of $q_{1}$ and $q_{2}$ is either $\left|\Phi^{+}\right\rangle $
or $\left|\Phi^{-}\right\rangle $; otherwise, the state is $\left|\Psi^{+}\right\rangle $
or $\left|\Psi^{-}\right\rangle $. TP2 can thus obtain partial information
about Alice's secret message. However, when TP1 sends the sequence
$S_{2}$, which includes $q_{2}$ of each EPR pair, to Bob, $S_{2}$
also contains TP1's decoy photons, where the positions and bases of
these decoy photons are unknown to TP2. If TP2's first CNOT operation
is performed on an X-basis decoy photon, for example, $q_{d}=\left|+\right\rangle $,
the result is as follows:

\begin{equation}
CNOT_{de}\left|+\right\rangle \otimes\left|0\right\rangle =\frac{1}{\sqrt{2}}\left(\left|++\right\rangle +\left|--\right\rangle \right)_{de}
\end{equation}

It can be seen that if Bob measures the decoy photon in X basis, the
measurement result will be $\left|+\right\rangle $ or $\left|-\right\rangle $
with an equal probability of $50\%$. Hence, if the decoy photon is
in $X$ basis, TP2's attack may disturb the state of decoy photon.
Eventually, it causes TP2 to be detected with a probability of $50\%$.
However, if the decoy photon is in Z basis, TP2's first CNOT operation
will not disturb the state. Assume that TP1 selects the basis of each
decoy photon with equal probability in Z basis or X basis. TP2's attack
will be detected with the following probability: $50\%\times50\%+50\%\times0=25\%$.
Consequently, if there are $n$ decoy photons, the detection rate
of TP2's attack is $1-\left(75\%\right)^{n}$. If $n$ is large enough,
the probability will be close to 1.

\subsubsection*{The Dense Coding Attack}

The external attacker, Eve, may try to perform the dense coding attack
\cite{DCA} to reveal Alice's secret message. When TP1 transmits $S_{1}$
to Alice in Step 2, Eve intercepts it, and prepares a sequence of
EPR states $\left|\Phi^{+}\right\rangle _{e1,e2}$, where $e1$ and
$e2$ respectively denote the first and the second particles of the
EPR states generated by Eve. Eve sends all $q_{e1}$, the first particle
of each EPR state, to Alice in hope that she successfully passes the
eavesdropping detection, and thus Alice's encoding operations will
be performed on Eve's $q_{e1}$. Consequently, when Alice sends out
the encoded qubits to TP2 in Step 5, Eve can retrieve her $q_{e1}$
and performs EPR measurement on every pair of $q_{e1}$ and $q_{e2}$.
That is, according to the measurement results (see Eq. (2),) Eve can
reveal Alice's secret message. However, $S_{1}$ contains decoy photons.
According to Eq. (1,) it can be seen that the first particle has two
measurement results in both two basis. If the original decoy photon
is $\left|1\right\rangle $, and Alice measures the fake photon, $q_{e1}$,
in Z basis, then the measurement result will be $\left|0\right\rangle $
or $\left|1\right\rangle $ with equal probability. If the measurement
result is $\left|0\right\rangle $, Eve's attack will be detected.
For each decoy photon, Alice will get an illegal measurement result
on Eve's fake photon with the probability of $50\%$. Let $n$ be
the number of decoy photons, Eve's attack will be detected with the
probability of $1-\left(50\%\right)^{n}$. If n is large enough, the
probability will be close to 1.

The above analyses denote that the proposed protocol is not only secure
against the general attack, but also secure against some special attacks.
If TP1 attacks the protocol, he/she will be detected in the public
discussion held by TP2 in Step 3. Similarly, if TP2 attacks the protocol,
he/she will be detected in Step 2, the public discussion of the decoy
photons generated by TP1. Two almost dishonest TPs, TP1 and TP2, share
duty to watch each other and as a result, two strangers, Alice and
Bob can have a secure communication between each other.

\subsubsection*{The Modification Attack}

Eve (TP1) may perform random unitary operations on the encoded qubits
when these qubits are sent from Alice to Bob via TP2. Hence Alice's
message could be modified \cite{Modification-1,Modification-2}. However,
the decoy photons are inserted in the quantum transmission, and Eve
does not know the positions of the decoy photons. Eve's random operations
will cause she being detected in the public discussion between Alice
and TP2 (or TP2 and Bob.) 

Considering the following situations: (1) Eve (TP1) might perform
only one operation in hope that the selected position is the encoded
qubit rather than the decoy photon; (2) TP2 performs the modification
attack, where TP2 knows all the positions of the decoy photons. Alice
and Bob can simplify use the message authenticate code to protect
the integrity of the transmitted secret message. The modification
thus can be detected.

\subsubsection*{The Deng-Long Criteria}

The Deng-Long criteria \cite{EMA-1} defines the requirements for
a secure quantum communication protocol. The requirements are listed
as follows:
\begin{enumerate}
\item A QSDC protocol does not require any additional classical information
transmissions except for the eavesdropping check. The receiver can
directly read the secret information after the quantum transmissions.
\item Eve, an eavesdropper, cannot obtain any useful information about the
secret message.
\item The sender and the receiver can detect Eve before they encode the
secret message on the quantum states.
\item The quantum states are transmitted in a block by block way.
\end{enumerate}
The following analyses respectively indicate the proposed QSDC protocol
satisfies the Deng-Long criteria.
\begin{enumerate}
\item In the proposed QSDC protocol, Bob can directly reveal Alice's secret
message according to his measurement results (see Step 7.) Alice sends
classical information in Step 2, Step 3, and Step 5, and they are
all for detecting the eavesdropping. Alice does not send any classical
information except for the eavesdropping check.
\item As shown in the above security analyses, Eve cannot obtain the secret
information sent by Alice.
\item In the proposed QSDC protocol, if Eve (TP1, or TP2) performs attacks
in the entanglement establishment process, Alice and Bob can detect
the attacks in the public discuss process in Step 2 and Step 3. After
confirming the security of the quantum transmission, Alice will encode
her message in Step 5.
\item The quantum transmissions in the proposed schemes, i.e., the entanglement
establishment and the QSDC, are sending a sequence of particles including
all the first (second) qubits of the EPR states, which is the ``block
by block way.''
\end{enumerate}

\section{Conclusions}

This paper presents a new method in quantum cryptography that allows
multiple strangers to establish an entanglement with the help of two
almost dishonest TPs. Each TP is designed to prevent the other TP
from acting maliciously; hence, both TPs can be almost dishonest.
The proposed protocol can also be easily transformed into a quantum
communication, a quantum teleportation, a quantum key distribution,
a quantum private comparison, etc., between two strangers. It is indeed
a challenging task to provide a scenario, secure entanglement establishment
between two strangers using other approaches. It would be an interesting
future research to have a secure entanglement establishment for strangers,
who cannot find two common TPs to help them.

\section*{Acknowledgment}

This research is partially supported by the Ministry of Science and
Technology, Taiwan, R.O.C., under the Contract No. MOST 104-2221-E-006-102-.

\section*{Appendix - Formal Security Model and Analysis}

In this section, we define the adversarial model of the two public
discussions in Step 2 and Step 3 of the proposed scheme. The security
of the first public discussion (Step 2) is analyzed in Section A.1
and then, the second public discussion (Step 3) is analyzed in Section
A.2.

\subsection*{A.1 The First Public Discussion}

In the following analyses, the public discussion between Alice and
TP1 is analyzed. Note that the security of the public discussion between
Bob and TP1 is the same as the one between Alice and TP1, hence we
omit that in the following description.

\textbf{Formal Security Model}

The security model of the interactions between an adversary and the
protocol participants occurs only via oracle queries which model the
adversary's capabilities in a real attack. Let $A$ denote Alice,
$TP1$ denote TP1, and $P1$ is the public discussion they participate.
The participants of $P1$ can launch more than one instance. Here
we allow a probabilistic polynomial time (PPT) adversary $\mathscr{A}$
to potentially control all the communication in the network via accessing
to a set of oracles as defined below. Let $A^{i}$ denotes the instance
$i$ of $A$. $TP1^{\text{j}}$ is the instance $j$ of $TP1$.
\begin{description}
\item [{Execute($A^{i}$,$TP1^{j}$):}] The query models the passive attack.
An adversary can obtain all messages exchanged between $A^{i}$ and
$TP1^{j}$.
\item [{Reveal($A^{i}$):}] In this query model, if the oracle has accepted,
it returns the secret quantum state between $A^{i}$ and $TP1^{j}$
to the adversary; otherwise, it returns the $null$ value to the adversary. 
\item [{Send($A^{i}/TP1^{j}$,$m$):}] This query models an active attack.
It returns the information corresponded to an input $m$ that $A^{i}$
or $TP1^{j}$ would send to each other.
\item [{Corrupt($A^{i}$,$a$):}] This query models corruption capability
of the adversary. If $a=0$, it returns a $null$ value; otherwise,
it returns the secret quantum states between $A^{i}$ and $TP1^{j}$.
\item [{Test($TP1^{j}$):}] This query measures whether the public discussion
is secure or not. By throwing an unbiased coin, $b$, if $b=1$, it
returns a random bit sequence with the same length as $A^{i}$'s measurement
result. The query can only be called once.
\end{description}
In this model, we consider two kinds of adversaries. A passive adversary
is allowed to issue the \textbf{Execute} and \textbf{Test} queries
and an active adversary is additionally allowed for sending the \textbf{Send}
query.

\textbf{Definitions of Security}

To demonstrate the security of the first public discussion, we will
give the security definition as follows.

\textbf{Definition 1} (Partnering): $A^{i}$ and $TP1^{j}$ are partnered,
if they mutually authenticate each other.

\textbf{Definition 2} (Freshness): An entity $A^{i}$ with the partner
$TP1^{j}$ is freshness if the following two conditions hold:

(1) If it has accepted an measurement result $MR\neq null$ and both
the entity and its partner have not been sent a \textbf{Reveal} query.

(2) There is no \textbf{Corrupt} query has been asked before the query
\textbf{Send} has been asked.

The advantage of the adversary $\mathscr{A}$ is measured by the ability
of distinguish a legal measurement result from a random value. We
define \textbf{Succ} to be an event that $\mathscr{A}$ correctly
guesses the bit \textbf{$b$}, which is chosen in the \textbf{Test}
query. Hence, the advantage of $\mathscr{A}$ in the attacked scheme
$P1$ is defined as: $Adv_{P1}\left(\mathscr{A}\right)=\left|2\times Pr\left[Succ\right]-1\right|$.
We argue that the public discussion $P1$ is secure, as $Adv_{P1}\left(\mathscr{A}\right)$
is negligible. Precisely, the adversary $\mathscr{A}$ does not have
any advantage to obtain the correct measurement result between the
participants. 

\textbf{Security Analysis}

In the following description, we show that the public discussion,
$P1$, holds several security properties, which are required for a
secure quantum cryptographic public discussion. Let the maximum advantage
of the adversary with running time $Tm$ be for a certain task denoted
as $Adv_{Task}\left(Tm\right)$. The following advantages will be
used in the analyses.

$Adv_{Qubit}^{Clone}\left(Tm\right)$: The advantage for cloning a
qubit.

$Adv_{A}^{Forge}\left(Tm\right)$: The advantage for impersonate himself/herself
as Alice ($A$). 
\begin{description}
\item [{Lemma1}] The advantage for cloning a qubit, $Adv_{Qubit}^{Clone}\left(Tm\right)$,
is negligible.
\item [{Proof}] The quantum no-cloning theory has already been well-proven
in several researches \cite{DECOY-3}, here, we briefly describe the
proof.\\
Assume that for an input qubit $q_{i}$ with an arbitrary state, there
exists a clone operation $U$. The clone operation can be defined
as follows:
\begin{equation}
\begin{array}{lll}
U\left|0\right\rangle _{i}\left|e\right\rangle _{o} & = & \left|0\right\rangle _{i}\left|0\right\rangle _{o}\\
U\left|1\right\rangle _{i}\left|e\right\rangle _{o} & = & \left|1\right\rangle _{i}\left|1\right\rangle _{o}\\
U\left|+\right\rangle _{i}\left|e\right\rangle _{o} & = & \left|+\right\rangle _{i}\left|+\right\rangle _{o},
\end{array}
\end{equation}
where $\left|e\right\rangle _{o}$ denotes the output qubit, and $\left|e\right\rangle $
is an arbitrary initial state.\\
Because $\left|+\right\rangle _{i}=\frac{1}{\sqrt{2}}\left(\left|0\right\rangle +\left|1\right\rangle \right)_{i}$,
it implies that $U\left|+\right\rangle _{i}\left|e\right\rangle _{o}=$
$\frac{1}{\sqrt{2}}\left(U\left|0\right\rangle _{i}\left|e\right\rangle _{o}+U\left|1\right\rangle _{i}\left|e\right\rangle _{o}\right)=$
$\frac{1}{\sqrt{2}}\left(\left|0\right\rangle _{i}\left|0\right\rangle _{o}+\left|1\right\rangle _{i}\left|1\right\rangle _{o}\right)$.
However, $U\left|+\right\rangle _{i}\left|e\right\rangle _{o}=$ $\left|+\right\rangle _{i}\left|+\right\rangle _{o}=$
\item [{$\frac{1}{\sqrt{2}}\left(\left|0\right\rangle _{i}\left|0\right\rangle _{o}+\left|0\right\rangle _{i}\left|1\right\rangle _{o}+\left|1\right\rangle _{i}\left|0\right\rangle _{o}+\left|1\right\rangle _{i}\left|1\right\rangle _{o}\right)$,}] which
is not equal to $\frac{1}{\sqrt{2}}\left(\left|0\right\rangle _{i}\left|0\right\rangle _{o}+\left|1\right\rangle _{i}\left|1\right\rangle _{o}\right)$.
The contradiction shows that the qubit cannot be cloned. $Adv_{Qubit}^{Clone}\left(Tm\right)$
is negligible.
\item [{Lemma2}] Suppose that there exists an attacker $\mathscr{A}$,
who impersonates as Alice ($A$) with the running time $Tm$ in the
public discussion. Then the advantage of $\mathscr{A}$, $Adv_{A}^{Forge}\left(Tm\right)=$$Adv{}_{Qubit}^{Clone}\left(Tm\right)$.
\item [{Proof}] Suppose that $\mathscr{A}$ impersonates as Alice. In Step
1 of the proposed scheme, TP1 sends a quantum sequence to Alice, and
discusses the decoy photons with Alice in Step 2. If $\mathscr{A}$
can successfully impersonate as Alice, then she can send her fake
photon to Alice, and TP1 cannot detect the problem.\\
When TP1 sends the qubit sequence $S_{1}$ to Alice, $\mathscr{A}$
constructs an attack $\beta$ to clone every qubit in $S_{1}$. The
sequence of the cloning outputs is denoted as $\hat{S_{1}}$. Then,
$\beta$ sends the original sequence $S_{1}$ to Alice. Alice will
acknowledge TP1 that she has received the qubits. Then TP1 will announce
the bases and positions of the decoy photons to Alice. Alice will
select the corresponding qubits from $S_{1}$ and measure them in
the bases TP1 announced. Alice then transmits all the measurement
results to TP1 and TP1 can compare the measurement results and his/her
initial states of decoy photons to detect the existence of the eavesdroppers.
Because these public classical informations are transmitted via the
authenticated channel shared between Alice and TP1, $\beta$ cannot
forge or modify them. Here, $\beta$'s goal is to successfully clone
the qubits from $S_{1}$ to $\hat{S_{1}}$. $\beta$ runs a subroutine
and simulates its attack environment, and gives all the required public
parameters to $\mathscr{A}$. Without losing the generality, assume
that $\mathscr{A}$ does not ask queries on the same message more
than once. $\beta$ maintains a list $L_{CloneQubit}$ to ensure identical
responding and avoid collision of the queries. $\beta$ simulates
the oracle queries of $\mathscr{A}$ as follows:

\begin{description}
\item [{Send-query:}] The send query is classified into the following types:\end{description}
\begin{itemize}
\item Send$\left(TP1^{j},S_{1}\right)$: $\beta$ clones every qubits in
the quantum sequence $S_{1}$, and forms the output qubits as a new
sequence $\hat{S_{1}}$. $\beta$ returns $\hat{S_{1}}$ to $\mathscr{A}$.
\item Send$\left(A^{i},ok\right)$: Alice sends the acknowledgement to TP1
for receiving qubits. $\beta$ direct pass the collected information
to $\mathscr{A}$.
\item Send$\left(TP1^{j},pos\&bases\right)$: TP1 announces the positions
and bases of the decoy photons to Alice. $\beta$ direct pass the
collected information to $\mathscr{A}$.
\item Send$\left(A^{i},mr\right)$: Alice sends the measurement results
to TP1. $\beta$ stores these results for the test query.\end{itemize}
\begin{description}
\item [{Execute-query:}] When $\mathscr{A}$ asks for an\textbf{ Execute($A^{i}$,$TP1^{j}$)}
query, $\beta$ returns the transcript $\left\langle \hat{S_{1}},\mbox{Send}\left(A^{i},ok\right),\mbox{Send}\left(TP1^{j},pos\&bases\right)\right\rangle $
to $\mathscr{A}$ by using the simulation of send query.
\item [{Test-query:}] When $\mathscr{A}$ makes the test query, if the
query is not asked in the first session, then $\beta$ will abort
it; otherwise, $\beta$ randomly chooses a bit $b$. If $b=0$, $\beta$
returns the value of Send$\left(A^{i},mr\right)$; otherwise, $\beta$
returns a random string to $\mathscr{A}$. The adversary has to distinguish
the random string from a legal measurement result. In order to do
that, if the quantum could be cloned, $\mathscr{A}$ can measure the
qubits from $\hat{S_{1}}$ by using the positions and bases obtained
from the query Send$\left(TP1^{j},pos\&bases\right)$. Then, the adversary
can successfully get the legal measurement results, hence the random
string and the legal measurement results can be distinguished. Hence,
the adversary's advantage, $Adv_{Alice}^{Forge}\left(Tm\right)=$$Adv{}_{Qubit}^{Clone}\left(Tm\right)$. 
\end{description}
\end{description}

\subsection*{A.2 The Second Public Discussion}

In the following analyses, the public discussion between Alice, Bob
and TP2 is analyzed.

\textbf{Formal Security Model}

Let $A$ denotes Alice, $B$ denotes Bob, $TP2$ denotes TP2, and
$P2$ is the public discussion they participate. To describe the multiple
instances of the participants, let $A^{i}$ denote the instance $i$
of $A$, $B^{j}$ denote the instance $j$ of Bob, and $TP2^{k}$
is the instance $k$ of $TP2$.
\begin{description}
\item [{Execute($A^{i}$,$B^{j}$,$TP2^{k}$):}] The query models the passive
attack. An adversary can obtain all messages exchanged between $A^{i}$,
$B^{j}$ and $TP2^{k}$.
\item [{Reveal($A^{i},B^{j}$):}] In this query model, if the oracle has
accepted, it returns the secret quantum state between $A^{i}$ and
$B^{j}$ to the adversary; otherwise, it returns the $null$ value
to the adversary. 
\item [{Send($A^{i}/B^{j}/TP2^{k}$,$m$):}] This query models an active
attack. It returns the information corresponded to an input $m$ that
$A^{i}$, $B^{j}$, or $TP2^{k}$ would send to each others.
\item [{Corrupt($A^{i}$,$B^{j}$,$a$):}] This query models corruption
capability of the adversary. If $a=0$, it returns a $null$ value;
otherwise, it returns the secret quantum states between $A^{i}$ and
$Bs{}^{j}$.
\item [{Test($TP2^{k}$):}] This query measures whether the public discussion
is secure or not. By throwing an unbiased coin, $b$, if $b=1$, it
returns a random bit sequence with the same length as $A^{i}$ and
$B^{j}$'s measurement results. The query can only be called once.
\end{description}
Similar as the previous model, we consider two kinds of adversaries.
A passive adversary is allowed to issue the \textbf{Execute} and \textbf{Test}
queries and an active adversary is additionally allowed for sending
the \textbf{Send} query.

\textbf{Definitions of Security}

To demonstrate the security of the first public discussion, we will
give the security definition as follows.

\textbf{Definition 1} (Partnering): $A^{i}$, $B^{j}$ and $TP2^{k}$
are partnered, if they mutually authenticate each other.

\textbf{Definition 2} (Freshness): The entities $A^{i}$ and $B^{j}$
with the partner $TP2^{k}$ is freshness if the following two conditions
hold:

(1) If it has accepted an measurement result $MR\neq null$ and both
the entity and its partner have not been sent a \textbf{Reveal} query.

(2) There is no \textbf{Corrupt} query has been asked before the query
\textbf{Send} has been asked.

The advantage of the adversary $\mathscr{A}$ is also measured by
the ability of distinguish a legal measurement result from a random
value. Hence, the advantage of $\mathscr{A}$ in the attacked scheme
$P2$ is defined as: $Adv_{P2}\left(\mathscr{A}\right)=\left|2\times Pr\left[Succ\right]-1\right|$.
We argue that the public discussion $P1$ is secure, as $Adv_{P2}\left(\mathscr{A}\right)$
is negligible. Precisely, the adversary $\mathscr{A}$ does not have
any advantage to obtain the correct measurement result between the
participants. 

\textbf{Security Analysis}

In the following description, we show that the public discussion,
$P2$, holds several security properties, which are required for a
secure quantum cryptographic public discussion. 

$Adv_{FakeState}^{Gen}\left(Tm\right)$: The advantage for generating
a fake entangled state without being detected in the second public
discussion.

$Adv_{P2}^{Attack}\left(Tm\right)$: The advantage for attacking $P2$
successfully.
\begin{description}
\item [{Lemma3}] The advantage for generating a fake entangled state $\left|\psi\right\rangle $
that can be written as $\left|\phi\right\rangle \left|\Phi^{+}\right\rangle $,
$Adv_{FakeState}^{Gen}\left(Tm\right)$ is negligible.
\item [{Proof}] Let $\left|\phi\right\rangle $ be an special entangled
state that an adversary can generate. The adversary will try share
this state with the legal users, Alice and Bob, before the second
public discussion. During the second public discussion, because Alice,
Bob, and TP2 will check if the state shared by Alice and Bob is $\left|\Phi^{+}\right\rangle $.
In this case, the adversary should make $\left|\phi\right\rangle =\left|\Phi^{+}\right\rangle \left|\psi\right\rangle $,
where $\left|\psi\right\rangle $ is the state held by the adversary,
and $\left|\Phi^{+}\right\rangle $ is shared among Alice and Bob.\\
However, $\left|\phi\right\rangle =$$\left|\Phi^{+}\right\rangle \left|\psi\right\rangle $
implies that $\left|\phi\right\rangle $ is a product state (i.e.,
it is the product of $\left|\psi\right\rangle $ and $\left|\Phi^{+}\right\rangle $),
which is not entangled. The contradiction shows that the advantage
to generate such special quantum state, $Adv_{FakeState}^{Gen}\left(Tm\right)$
is negligible.
\item [{Lemma4}] Suppose that there exists an attacker $\mathscr{A}$,
who wants to successfully attack $P2$ with running time $Tm$ in
the public discussion. Then $Adv_{P2}^{Attack}\left(Tm\right)=$$2\times Adv{}_{Qubit}^{Clone}\left(Tm\right)+$$Adv_{FakeState}^{Gen}\left(Tm\right)$.
\item [{Proof}] Suppose that $\mathscr{A}$ wants to attack $P2$ procedure.
The adversary hopes that he/she could share an entangled state with
Alice, Bob, and himself/herself. $\mathscr{A}$ constructs an attack
$\gamma$ to help him/her. $\gamma$ will generate a special quantum
state and distribute them to Alice, Bob, and the adversary. To send
fake qubits to Alice and Bob without being detected, $\gamma$ has
to pass the first public discussions between Alice and TP1 (Bob and
TP1.) Then, when the second public discussion is started, the fake
qubits held by Alice and Bob can be converted  to $\left|\Phi^{+}\right\rangle $,
then the second public discussion will be success.\\
Here, $\gamma$'s goal is to successfully generate a fake entangled
state, and pass the first public discussions. $\gamma$ runs a subroutine
and simulates its attack environment, and gives all the required public
parameters to $\mathscr{A}$. Without losing the generality, assume
that $\mathscr{A}$ does not ask queries on the same message more
than once. $\gamma$ maintains a list $L_{GenFakeState}$ to ensure
identical responding and avoid collision of the queries. $\gamma$
simulates the oracle queries of $\mathscr{A}$ as follows:

\begin{description}
\item [{Send-query:}] the send query is defined as follows:\end{description}
\begin{itemize}
\item Send$\left(TP1^{k},S_{1}/S_{2}\right)$: when $TP1^{k}$ sends $S_{1}$
($S_{2}$) to $A^{i}$ ($B^{j}$), $\gamma$ generates a sequence
of a n-qubit fake state $\left|\phi\right\rangle _{123...n}$, sends
all the first qubits $q_{1}$ to Alice and all the second qubit $q_{2}$
to Bob. The remained qubits $q_{3...n}$ of all the fake states are
denoted as $\hat{S_{3...n}}$ to $\mathscr{A}$.
\item Send$\left(A^{i}/B^{j},ok\right)$: Alice and Bob will notify TP2
the first public discussion has been success. $\gamma$ direct pass
the collected information to $\mathscr{A}$.
\item Send$\left(TP2^{k},pos\&bases\right)$: TP2 announces the positions
and bases to Alice and Bob. $\gamma$ direct pass the collected information
to $\mathscr{A}$.
\item Send$\left(A^{i}/B^{j},mr\right)$: Alice and Bob sends the measurement
results to TP1. $\gamma$ stores these results for the test query.\end{itemize}
\begin{description}
\item [{Execute-query:}] When $\mathscr{A}$ asks for an\textbf{ Execute($A^{i}$,$B^{j}$$TP2^{k}$)}
query, $\gamma$ returns the transcript $\left\langle \hat{S_{3...n}},\mbox{Send}\left(A^{i}/B^{j},ok\right),\mbox{Send}\left(TP2^{k},pos\&bases\right)\right\rangle $
to $\mathscr{A}$ by using the simulation of send query.
\item [{Test-query:}] When $\mathscr{A}$ makes the test query, if the
query is not asked in the first session, then $\gamma$ will abort
it; otherwise, $\gamma$ randomly chooses a bit $b$. If $b=0$, $\gamma$
returns the value of Send$\left(A^{i}/B^{j},mr\right)$; otherwise,
$\beta$ returns a random string to $\mathscr{A}$. The adversary
has to distinguish the random string from the legal measurement results.
In order to do that, if the special entangled state $\left|\phi\right\rangle _{123...n}$
can be converted to $\left|\phi\right\rangle _{123...n}=\left|\Phi^{+}\right\rangle _{12}\left|\psi\right\rangle _{3...n}$,
Alice and Bob can generate a legal pair of measurement results, and
$\mathscr{A}$ can obtain their measurement result from $\left|\psi\right\rangle _{3...n}$.
To success such attack, $\gamma$ has to impersonate as Alice and
Bob to respectively pass the two public discussions (i.e., the first
public discussion between Alice and TP1 and between Bob and TP1.)
Hence, the adversary's advantage can be derived as $Adv_{P2}^{Attack}\left(Tm\right)=$$Adv{}_{Alice}^{Forge}\left(Tm\right)+Adv{}_{Bob}^{Forge}\left(Tm\right)+Adv_{FakeState}^{Gen}\left(Tm\right)$.
According to Lemma2, $Adv_{P2}^{Attack}\left(Tm\right)=$$2\times Adv{}_{Qubit}^{Clone}\left(Tm\right)+Adv_{FakeState}^{Gen}\left(Tm\right)$.
\end{description}
\end{description}
\bibliographystyle{IEEEtran}
\bibliography{Untitled}

\end{document}